\documentclass[journal=jctcce,manuscript=article]{achemso}

\usepackage{multirow}
\usepackage{subfig}
\usepackage{graphicx}
\usepackage[version=3]{mhchem} 

\author{Anna Thorn Ekstrøm}
\affiliation{University of Copenhagen, Universitetsparken 5, DK-2100 Copenhagen Ø, Denmark}
\author{Stephan P. A. Sauer}
\affiliation{University of Copenhagen, Universitetsparken 5, DK-2100 Copenhagen Ø, Denmark}
\email{sauer@chem.ku.dk}

\title[Vibrational g-Factors of the Diatomic Molecules]
  {A Computational Study of the Vibrational and Rotational g-Factors of the Diatomic Molecules LiH, LiF, CO, CS, SiO and SiS$^\dag$}


\begin{document}







\begin{abstract}
The purpose of this article is to present theoretical values for the vibrational and rotational g-factors of several diatomic molecules.
The calculations have been carried out at the Multi-Configurational Self-Consistent Field (MCSCF) level of theory. To determine the most reliant method and basis set for these calculations also the Hartree-Fock (HF) and Density Functional Theory (DFT) approaches were considered. Different DFT functionals, including B3LYP, BHandHLYP, PBE0, B3PW91 and KT3 have been employed. Furthermore, different active spaces were evaluated to optimize MCSCF. To establish the accuracy of the methods the computed rotational g-factors were compared to experimental values. The benchmark study of CO and CS shows that the MCSCF method provides most reliable results and that the aug-cc-pCV5Z basis set is the most sufficient. The aug-cc-pCVQZ basis set for Li and aug-cc-pV5Z basis set for H gave best results for LiH. The active spaces tested for CO and CS do not yet converge towards the experimental values when more determinants were included. However, if the g-factors are vibrationally averaged, the computed values are seen to move towards the experimental value. 
Lastly, the g-factors have been vibrationally averaged, and it shows a slight improved agreement between computed values and experimental data.
\end{abstract}

\section{Introduction}
The Born-Oppenheimer (BO) approximation is instrumental in solving the non-relativistic Schrödinger equation and predicting molecular properties. It assumes that the nuclei, due to a heavier mass compared to electrons, can be considered stationary while treating the electrons and that the electrons will adjust instantaneously to a change in the position of the nuclei. This allows for the separation of nuclear and electronic motion. Within the Born-Oppenheimer approximation, the electronic energy can then be obtained by solving the electronic Schrödinger equation for a fixed position of the nuclei. 
The BO approximation is generally considered to be a good approximation. However, it breaks down when two solutions of the electronic Schrödinger equation are energetically close.\cite{Comp_chem}
To obtain a more precise description of molecules, the consequences of lifting the restrictions of the BO approximation are investigated by allowing the movement of the nuclei to influence the electrons. The coupling between electronic and nuclear motion gives rise to an adiabatic and a non-adiabatic contribution.
In 1966, Herman and Asgharian derived an effective Hamiltonian and identified adiabatic and non-adiabatic corrections.\cite{herman1966theory} They showed that the non-adiabatic correction to the rotational reduced mass, $\alpha (R)$, is proportional to the rotational g-factor and interpreted the rotational non-adiabatic contribution to be a consequence of the electrons not rotating perfectly with the nuclei. While the rotational g-factor can be measured through rotational Zeeman or molecular beam experiments,\cite{spas035} there is no equivalent relationship to an physical observable for the non-adiabatic contribution to the vibrational reduced mass, $\beta (R)$. 
Nevertheless, Herman and Ogilvie defined in 1998 a similar vibrational g-factor proportional to the non-adiabatic vibrational correction.\cite{herman1998effective}
Bunker and Moss derived in 1977 an effective Hamiltonian that included non-adiabatic corrections, which they applied to $\text{H}_2$ and $\text{D}_2$.\cite{bunker1977breakdown, bunker1977application} In 1980, they extended the effective Hamiltonian to triatomic molecules.\cite{bunker1980effect} Using the Bunker and Moss formalism, Schwenke computed ab initio corrections to the Born-Oppenheimer approximation in 2001.\cite{schwenke2001first, schwenke2001beyond} He reported accurate results for both one-electron systems $\text{H}_2^+$ and $\text{HD}^+$ and multi-electron systems $\text{H}_2$ and $\text{H}_2\text{O}$ using several different correlation consistent basis set. His article concludes that the basis sets with diffuse functions are important to obtain precise non-adiabatic corrections. Furthermore, the influence of the adiabatic and non-adiabatic correction terms were discussed.

Recently, computing the non-adiabatic corrections in order to achieve better agreement with experimental values has received renewed attention. In the most recent work by Koput in 2024, he discussed adiabatic and non-adiabatic effects and the accurate potential energy for the diatomic molecules CO and MgH.\cite{koput2024toward, KoputMgH} Koput predicted the potential energy for the CO molecule in its ground state using the Coupled Cluster approach up to the CCSDTQP level employing the aug-cc-pCVXZ basis set with X going to octuple-zeta level. The adiabatic energy correction was computed using the CCSD/aug-cc-pCVQZ level of theory, and the effects beyond the adiabatic approximation were accounted for through calculating the g-factors using the complete active space self-consistent field (CASSCF) method employing the aug-cc-pV5Z basis set. The non-adiabatic corrections to the reduced masses were calculated to be $\alpha = -0.41\times 10^{-3}$ and $\beta = -0.25\times 10^{-3}$. The vibrational averaged g-factors were predicted to be $\left<g_r\right>=-0.259$ and $\left<g_v\right>=0.046$.
For MgH, a radical, he used the multireference averaged coupled-pair functional (MRACPF) method and the aug-cc-pCVXZ level of quality with X=5-8. To compute the g-factors the CASSCF method with the aug-cc-pV5Z basis set was employed again. The non-adiabatic parameters were reported as $\alpha = -1.40\times 10^{-3}$ and $\beta = -0.67\times 10^{-3}$. He concluded that the vibration-rotation energy levels are close to spectroscopic accuracy. 
Through his work, Koput confirmed that including adiabatic and non-adiabatic effects leads to a significant improvement of spectroscopic constants compared to experimental data. This emphasizes the importance of considering the adiabatic and non-adiabatic contributions.

Previous work by our group has investigated the rotational and vibrational g-factors for multiple diatomic molecules including H$_2$, LiH, BH, AlH, GaH, LiF, BF, AlF, GaF, CH$^+$, CO, SiO, GeO, HF, HCl, HBr, HI, HeH$^+$, NeH$^+$, KrH$^+$., ArH$^+$ and XeH$^+$.\cite{spas002, spas003, sauer1994experimental, spas015, sauer1996theoretical, sauer1998relation, enevoldsen2001relativistic, bak2005vibrational, sauer2005quantum, spas067, kjaer2009relation, spas096, sauer2011calculated} The non-adiabatic contributions were computed with the Dalton program\cite{dalton, spas191} using the self-consistent field or multi-configurational self-consistent field methods.
Rotational g-factors alone have been studied many times using many different electronic structure theory methods.\cite{spas035} Recent work on some of the molecules studied in this work include a DFT study using the aug-cc-pVTZ basis set on CO and CS\cite{mag06-jctc2-827} and DFT and CCSD(T) calculations using the aug-cc-pCVQZ basis set for LiH, LiF and CO.\cite{mag09-jcp131-144104}

As the previous literature states, the g-factors are essential for predicting accurate spectroscopic values and understanding the g-factors provides a deeper insight into the vibration-rotation energy levels in a molecule. Although rotational g-factors can be measured experimentally, computational studies remain the only method to assess the vibrational g-factor.
It is crucial to be able to compute accurate g-factors to achieve high-precision spectroscopic data. This requires a careful selection of computational methods. Hence, it becomes critical to determine the most efficient conditions for different computational methods. This involves choosing a suitable functional in DFT or selecting an appropriate active space in MCSCF calculations. Additionally, it is relevant to investigate the influence of basis sets when computing g-factors.

In the present study we will thus present calculated values for the vibrational and rotational g-factors of LiH, LiF, CO, CS, SiO and SiS. 
To our knowledge, these will be the first values of the vibrational g-factor reported for CS and SiS, while we present improved values for LiH, LiF, CO and SiO due to a more careful selection of the basis set. Furthermore, vibrationally averaged values of the g-factors will be presented for the first time for LiH, LiF, CS, SiO and SiS.

\section{Theoretical background}
The analysis of adiabatic and non-adiabatic contributions for an electronic state $\left|\Psi_0\right>$ of symmetry $^1\Sigma^+$ is based on an effective vibration-rotational Hamiltonian that can be derived through a Van Vleck transformation.\cite{Van_Vleck_Brion} In the Van Vleck transformation the effective Hamiltonian is defined as a sum over matrix elements from different perturbation orders of a transformed Hamiltonian,
\begin{equation}
    \Hat{H}^{\text{eff}} = \left<0|\Hat{\Tilde{H}}|0\right> =  \left<0|\Hat{\Tilde{H}}^{(0)}|0\right> + \lambda \left<0|\hat{\Tilde{H}}^{(1)}|0\right> + \lambda^2 \left<0|\Hat{\Tilde{H}}^{(2)}|0\right> + \cdots.
\end{equation}
Evaluating the matrix elements individually reveals that the adiabatic contribution simply becomes an expectation value of the considered electronic state. The non-adiabatic contributions arise from coupling with different electronic states. Following the Bunker and Moss formalism,\cite{bunker1977application} the evaluated effective Hamiltonian becomes
\begin{eqnarray}
\hat{H}^{\text{eff}}\!\!\!\!\!\!&=&\!\!\!\!\!\!-\frac{\hbar^2}{2} \frac{\partial}{\partial R} \frac{1}{\mu} \left(1+\frac{m_e}{m_p}g_v(R)\right)\frac{\partial}{\partial R} + \frac{1}{2 \mu R^2} \left(1+\frac{m_e}{m_p}g_r(R)\right) \hat{\Vec{J}}^2\nonumber\\ 
&&+E_0^{(0)}(R)+E^{\text{ad}}(R)+E^{\text{nad}}(R).
\end{eqnarray}
It contains corrections to the potential energy and corrections to the vibrational and rotational reduced masses. The energy consists of the potential energy within the BO approximation, $E_0^{(0)}=\left<\Psi_0|\hat{H}^{(0)}|\Psi_0\right>$, and additionally two energy contributions; the adiabatic contribution, $E^{\text{ad}}(R)$, and the non-adiabatic contribution $E^{\text{nad}}(R)$. 

The rotational and vibrational g-factors, $g_r(R)$ and $g_v(R)$ contribute to the reduced masses of the rotational or vibrational motions. 
The electronic contributions to the two g-factors calculated with nuclear masses, $g^{\text{el}}_{r,n}(R)$ and $g^{\text{el}}_{v,n} (R)$, are directly proportional to the non-adiabatic Born-Oppenheimer breakdown (BOB) parameters, $\alpha (R)$ and $\beta (R)$.
\begin{equation}\label{eq: alpha1}
    g^{\text{el}}_{r,n} (R)=\frac{m_p}{m_e} \alpha(R)
\end{equation}
and
\begin{equation} \label{eq: beta1}
    g_{v,n}^{\text{el}}(R) = \frac{m_p}{m_e} \beta(R).
\end{equation}
The BOB parameters $\alpha (R)$ and $\beta (R)$ account for the contribution of the electron masses to the reduced masses for the vibrational or rotational kinetic energy. The adiabatic and non-adiabatic corrections are direct consequences of electrons failing to follow the nuclei perfectly. They can be expressed as sum-over-states expressions.
The g-factors are a sum of a nuclear and an electronic contribution. 
The nuclear contribution is constant for a diatomic molecule and depends on atomic masses and charges of the molecules. The electronic contributions are determined through response functions. 
\begin{equation}\label{eq: gv}
    g_v = m_p\frac{Z_Am_B^2+Z_Bm_A^2}{(m_A+m_B)m_Am_B} - \frac{m_p}{m_e}\frac{\hbar^2}{\mu}\left<\left<\frac{\partial}{\partial R}; \frac{\partial}{\partial R} \right>\right>_{\omega=0}
\end{equation}
and
\begin{equation}
    g_r = m_p\frac{Z_Am_B^2+Z_Bm_A^2}{(m_A+m_B)m_Am_B} + \frac{m_p}{m_e}\frac{1}{\mu R^2}\left<\left< \Hat{L}_\bot; \Hat{L}_\bot \right>\right>_{\omega=0}.
\end{equation}
The nuclear part is always positive and the electronic part is inevitably negative, thus the overall sign of the g-factor is determined by the magnitudes of the two contributions. 

Partitioning the g-factors into two isotopically independent components is beneficial for fitting vibration-rotation spectra. This was achieved by Watson by defining two canonical momentum operators, one for atom A and one for atom B\cite{watson1973isotope, watson1980isotope, kjaer2009relation}
\begin{eqnarray}\label{eq: Isotoically invariant operators}
    \hat{P}_{zA/B}=\hat{P}_R + \frac{\left(R_{B/A,z}-R_{CM,z}\right)}{R} \sum_i \hat{p}_{i,z},
\end{eqnarray}
where $z$ indicates that the molecule is aligned along the $z$-axis. The isotopically independent g-factors become then
\begin{eqnarray}\label{eq gvA_respons}
    g_v^A=\frac{m_p}{\mu}Z_B + \frac{m_p}{m_e}\frac{1}{\mu}\left<\left<\hat{P}_{zA}; \hat{P}_{zA} \right>\right>_{\omega=0} 
\end{eqnarray}
and
\begin{eqnarray}\label{eq gvB_respons}
    g_v^B=\frac{m_p}{\mu}Z_A + \frac{m_p}{m_e}\frac{1}{\mu}\left<\left<\hat{P}_{zB}; \hat{P}_{zB} \right>\right>_{\omega=0}.
\end{eqnarray}
When considering one of the two atoms as the origin instead of the center of mass, one can separate the dependence of mass of the vibrational g-factor into two terms. 
The electronic matrix element of $\left<\psi_0|\hat{P}_{zA/B}|\psi_n\right>$ can be separated into five separate terms. Assuming that the molecule is aligned along the $z$-axis, the positions can be rephrased in terms of reduced mass through $\frac{R_{A,z}-R_{CM,z}}{R} = -\frac{m_B}{m_A+m_B} = -\frac{\mu}{m_A}$. Each term can be evaluated individually and rewritten by invoking the hypervirial theorem and resolution of identity. Thus, a new expression for the isotopically independent g-factors becomes
\begin{eqnarray}\label{eq: gA dipole}
    g_v^A=g_v+\frac{2m_p}{em_A}\frac{\partial}{\partial R}(\mu_z(\Vec{R}_{CM},R))+\frac{\mu m_p}{m_A^2} Q
\end{eqnarray}
and 
\begin{eqnarray} \label{eq: gB dipole}
    g_v^B=g_v-\frac{2m_p}{em_B}\frac{\partial}{\partial R}(\mu_z(\Vec{R}_{CM},R))+\frac{\mu m_p}{m_B^2} Q.
\end{eqnarray}
Combining these equations yields an expression for the total g-factor for a neutral molecule ($Q=0$)
\begin{eqnarray} \label{eq: g_V_3_cont}
    g_v = \frac{1}{2}\left\{g_v^A+g_v^B\right\}-\frac{m_p}{e}\frac{\partial}{\partial R}\left\{\mu_z(\Vec{R}_{CM},R)\right\}\left(\frac{1}{m_A}+\frac{1}{m_B}\right).
\end{eqnarray}
This expression shows that three factors influence the total vibrational g-factor: an irreducible non-adiabatic correction, which includes the partitioned vibrational g-factors, along with a contribution from the gradient of the dipole moment with respect to the internuclear distance, $R$. Furthermore, for non-neutral molecules, the total charge of the molecule also contributes through the last terms of equations \eqref{eq: gA dipole} and \eqref{eq: gB dipole}.\cite{kjaer2009relation}

\section{Computational Details}
We utilized the Dalton program to compute the electronic  contributions to the g-factors.\cite{dalton, spas191} 
The electronic energy and electronic contributions to the g-factors were obtained using the Hartree-Fock (HF) or self-consistent field (SCF) method, the multi-configurational self-consistent field (MCSCF) method and several density functional theory (DFT) methods. All MCSCF calculations were of the complete active space (CASSCF) type.\cite{roos2007complete}
The active spaces were chosen based on the occupation numbers of MP2 natural orbitals.\cite{JCP-88-3824-1988-Jensen, JCP-89-5354-1988-Jensen} 
Furthermore, for DFT calculations different exchange-correlation functionals have been employed to investigate if the DFT method can provide as reliable results as the MCSCF method, and to determine what functionals among the hybrid functionals, B3LYP\cite{becke1988density, miehlich1989results}, BHandHLYP\cite{becke1993new}, B3PW91\cite{becke1988density, perdew1993erratum}, PBE0\cite{ernzerhof1999assessment, adamo1999toward} and the generalized gradient approximated (GGA) functional KT3\cite{keal2004semiempirical} are the most optimal in terms of computing g-factors. 

In the basis set study, the investigation of the active spaces and the method comparison the calculations were carried out at the experimental bond lengths as given in Table \ref{tab: Dis}.
\begin{table}[h!]
    \caption{Equilibrium distances in Å obtained with HF and MCSCF calculations and experimentally determined values.}
    \label{tab: Dis}
    \centering
    \begin{tabular}{l l c c c}
    \hline
Molecule & active space & HF & MCSCF & Exp. \\ \hline
CO  & (10e,6o)  &\multirow{3}{3em}{1.1020}&1.1330 &\multirow{3}{3em}{1.1283}\\
    & (10e,10o) &                         &1.1290&\\
    & (6e,20o)  &                         &1.1270&\\
CS  & (10e,8o)  &\multirow{2}{3em}{1.5090}&1.5450&\multirow{2}{3em}{1.5350}\\
    & (10e,13o) &                         &1.5440&\\
SiO & (10e,8o)  &1.4750&1.5170&1.5097\\
SiS & (10e,13o) &1.9080&1.9370&1.9292\\
LiH & (4e,5o)   &1.6060&1.6110&1.5957\\
LiF & (8e,8o)   &1.5540&1.5800&1.5639\\
    \hline
    \end{tabular}
\end{table}
The partitioned g-factor expressions, represented by eqs. \eqref{eq: gA dipole}, \eqref{eq: gB dipole} and \eqref{eq: g_V_3_cont}, are only fulfilled when a complete one-electron basis set is employed. Consequently, achieving agreement between eqs. \eqref{eq: gv} and \eqref{eq: g_V_3_cont} confirms that the basis set is adequate. The discrepancy when comparing the equations stems solely from basis set limitations, since the hypervirial and resolution of the identity relations are fulfilled for variational wave functions, such as HF, MCSCF and Kohn-Sham DFT in the limit of a complete one-electron basis set. 
In a previous study the aug-cc-pVQZ basis set was found to be sufficient for computing the g-factors of different isotopic variants of $\text{H}_2$.\cite{bak2005vibrational} 
Going beyond the aug-cc-pVQZ basis set did not provide any significant change in the values of the g-factor of $\text{H}_2$. 
In a CCSD(T) study of rotational g-factors for small molecules including LiH, LiF and CO a large basis set study was also carried out concluding that the aug-cc-pCVQZ basis set would be sufficient.\cite{mag09-jcp131-144104}
Most recently, Koput employed the aug-cc-pV5Z basis set in his study of rotational and vibrational g-factors of CO.\cite{koput2024toward}
The current study focuses now on determining the g-factors for more, heteronuclear diatomic molecules and which basis sets are necessary for obtaining converged results. 
The g-factors computed from eqs. \eqref{eq gvA_respons} and \eqref{eq gvB_respons} using various Dunning type basis sets are therefore benchmarked against the g-factors achieved with eqs. \eqref{eq: gA dipole} and \eqref{eq: gB dipole}, to evaluate the basis set completeness. Several types of correlation consistent basis sets, cc-pVXZ, aug-cc-pVXZ, d-aug-cc-pVXZ and aug-cc-pCVXZ, are employed with X being D, T, Q or 5.\cite{kendall1992electron}

\begin{figure}[h!]
 \centering
 \includegraphics[width=15cm]{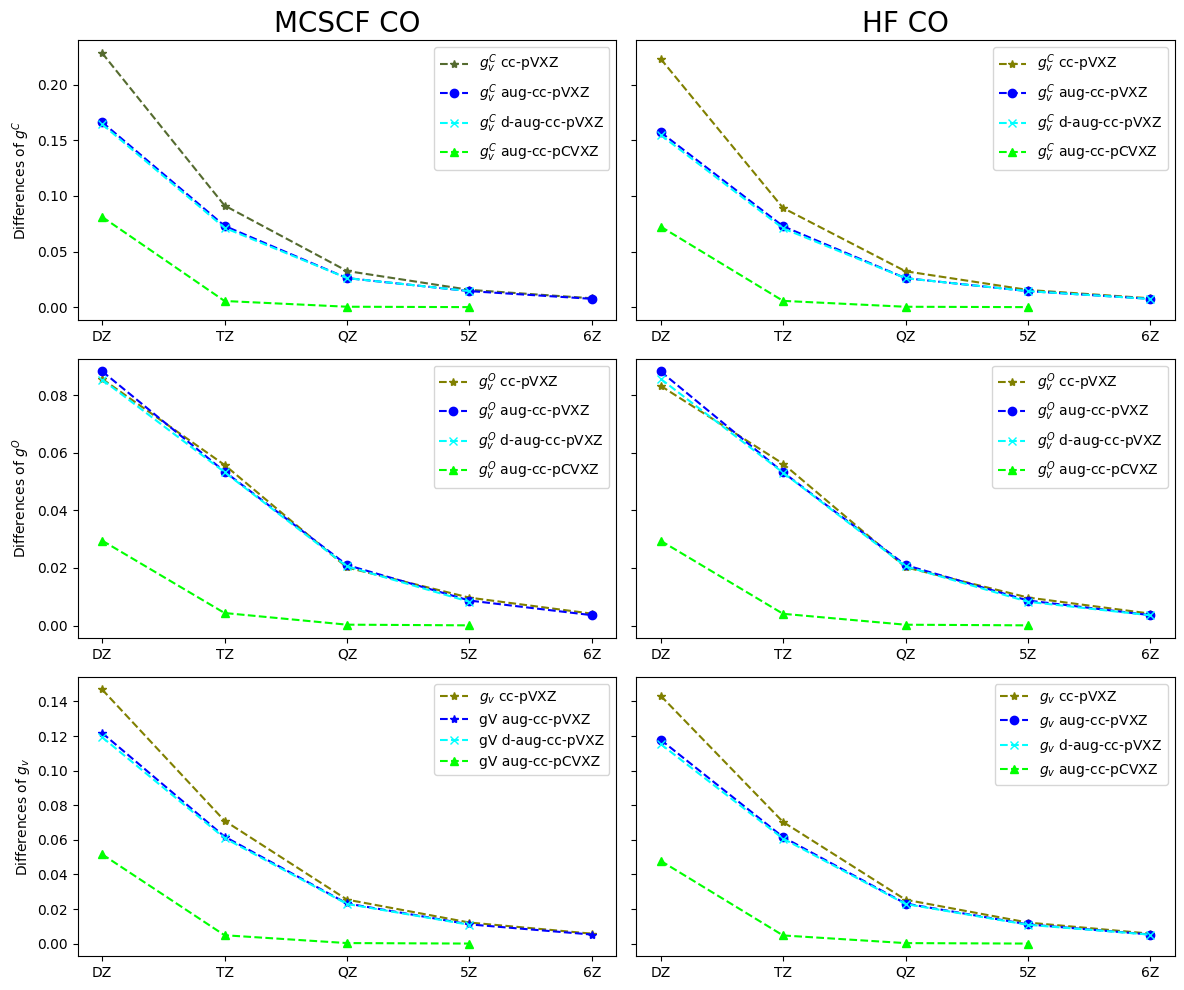}
 \caption{CO: Differences in the isotopically independent vibrational g-factors, $g_v^C$ and $g_v^O$, and the total vibrational g-factor, $g_v$, calculated from eqs. \eqref{eq gvA_respons}, \eqref{eq gvB_respons} and \eqref{eq: gv} versus from eqs. \eqref{eq: gA dipole}, \eqref{eq: gB dipole} and \eqref{eq: g_V_3_cont} as a function of the basis set at the MCSCF and HF level of theory. The active space is (10e, 17o).}
    \label{fig: CO_basisset}
\end{figure}
\section{Results and discussion}
\section{Results and Discussion}
In the following we will present our results for the vibrational and rotational g-factors of LiH, LiF, CO, CS, SiO and SiS and discuss their geometry dependence and thus the effect of vibrational averaging. But first, we will study the dependence of the calculated values on the choice of basis set and computational method focusing in particular on the choice of exchange correlation functional in the DFT calculations and of the active space in the MCSCF calculation.
\subsection{Selection of Basis Set}
We will start with the choice of the basis set and study it for the molecules CO, CS and LiH. The derivation of the partitioned g-factors, \eqref{eq: gA dipole}, \eqref{eq: gB dipole} and \eqref{eq: g_V_3_cont}, relies on the resolution of identity and the hypervirial theorem, and is thus strictly only fulfilled for variational methods in a complete basis. 

The sufficiency of the basis set for the calculation of the vibrational g-factors is in the following thus investigated by comparing the results obtained from eqs. \eqref{eq: gA dipole}, \eqref{eq: gB dipole} and \eqref{eq: g_V_3_cont} with the corresponding results from eqs. \eqref{eq gvA_respons}, \eqref{eq gvB_respons} and \eqref{eq: gv}. Good agreement ensures basis sets, which offer high reliability in the calculated g-factors.

For CO, fig. \ref{fig: CO_basisset} shows the difference between the $g_v^C$ values, the $g_v^O$ values, and the $g_v$ values for MCSCF (left) and HF (right) calculations. For both methods, the aug-cc-pVXZ and d-aug-cc-pVXZ basis sets provide equally accurate results. This indicates that inclusion of the second set of diffuse functions is unnecessary. For $g_v^O$ even the augmentation with diffuse functions can be neglected, since cc-pVXZ offers equivalent precision relative to the aug-cc-pVXZ and d-aug-cc-pVXZ basis sets throughout all zeta levels. 
Across all basis sets, it is evident that the sufficiency of the basis sets improves as the zeta level increases. For quadruple and higher zeta levels, the d-aug-cc-pVXZ, aug-cc-pVXZ, and cc-pVXZ ensure equal results.
However, including extra core-valence correlation functions in the aug-cc-pCVXZ basis set greatly improves the results for all parameters. This basis set consistently shows the fastest convergence and yields the most accurate results. 

\begin{figure}[h!]
 \centering
 \includegraphics[width=15cm]{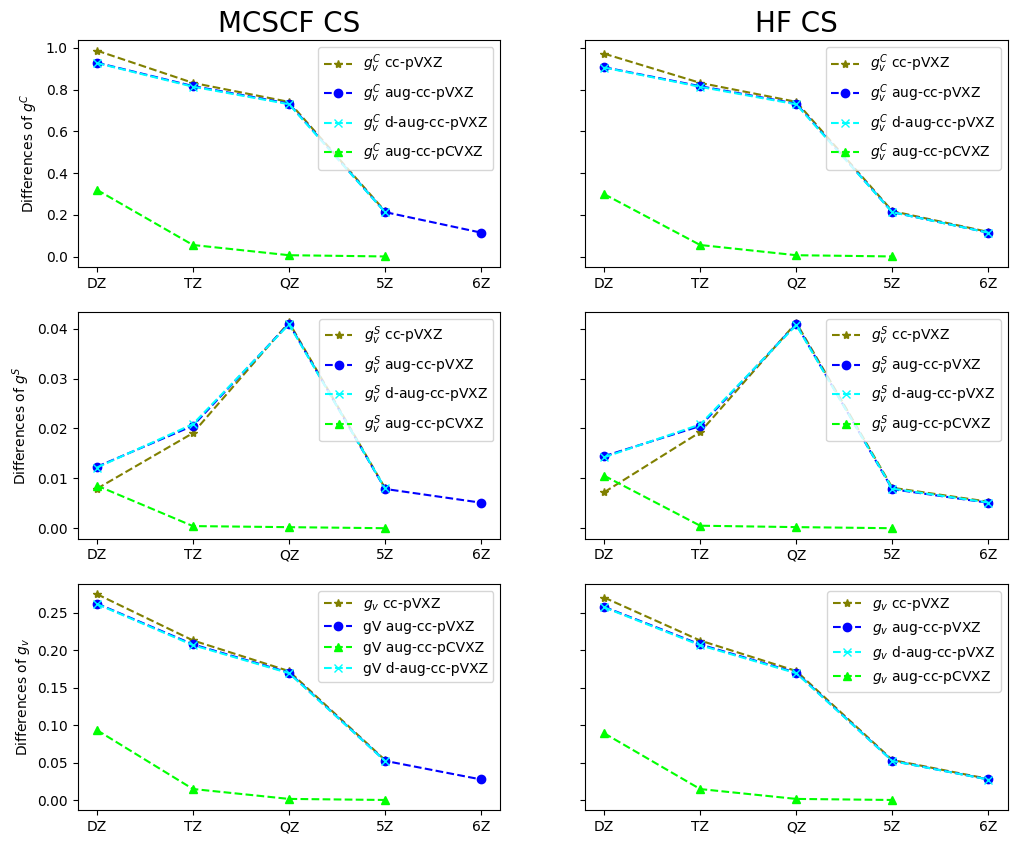}
 \caption{CS: Dfferences in the isotopically independent vibrational g-factors, $g_v^C$ and $g_v^S$, and the total vibrational g-factor, $g_v$, calculated from eqs. \eqref{eq gvA_respons}, \eqref{eq gvB_respons} and \eqref{eq: gv} versus from eqs. \eqref{eq: gA dipole}, \eqref{eq: gB dipole} and \eqref{eq: g_V_3_cont} as a function of the basis set at the MCSCF and HF level of theory. The active space is (10e, 16o).}
    \label{fig: CS_basisset}
\end{figure}
A similar analysis has been performed for CS as shown in Fig. \ref{fig: CS_basisset}. 
It mirrors the conclusions of CO although the behaviour of $g_v^S$ differs significantly from the behaviour of $g_v^O$. 
When increasing the zeta level the agreement between the two formulations becomes consistently better for $g_v^C$ and $g_v$. The largest improvement is seen on going from QZ to 5Z. Furthermore, the inclusion of extra core-valence correlation functions again dramatically enhances the convergence of the results at all zeta levels.

When going from DZ to QZ an unexpected trend is observed for $g_v^S$. The differences enlarge noticeably for the d-aug-cc-pVXZ, aug-cc-pVXZ, and cc-pVXZ basis set. The same trend is not observed for aug-cc-pCVXZ. This empathizes the importance of the inclusion of extra core-valence functions and furthermore, it could suggest a problem with the QZ basis set for sulfur. It appears that the d-aug-cc-pVXZ, aug-cc-pVXZ, and cc-pVXZ basis set all yield nearly identical results for all zeta levels apart from DZ, where the cc-pVXZ deviates slightly.

\begin{figure}[h!]
    \centering
    {{\includegraphics[width=15cm]{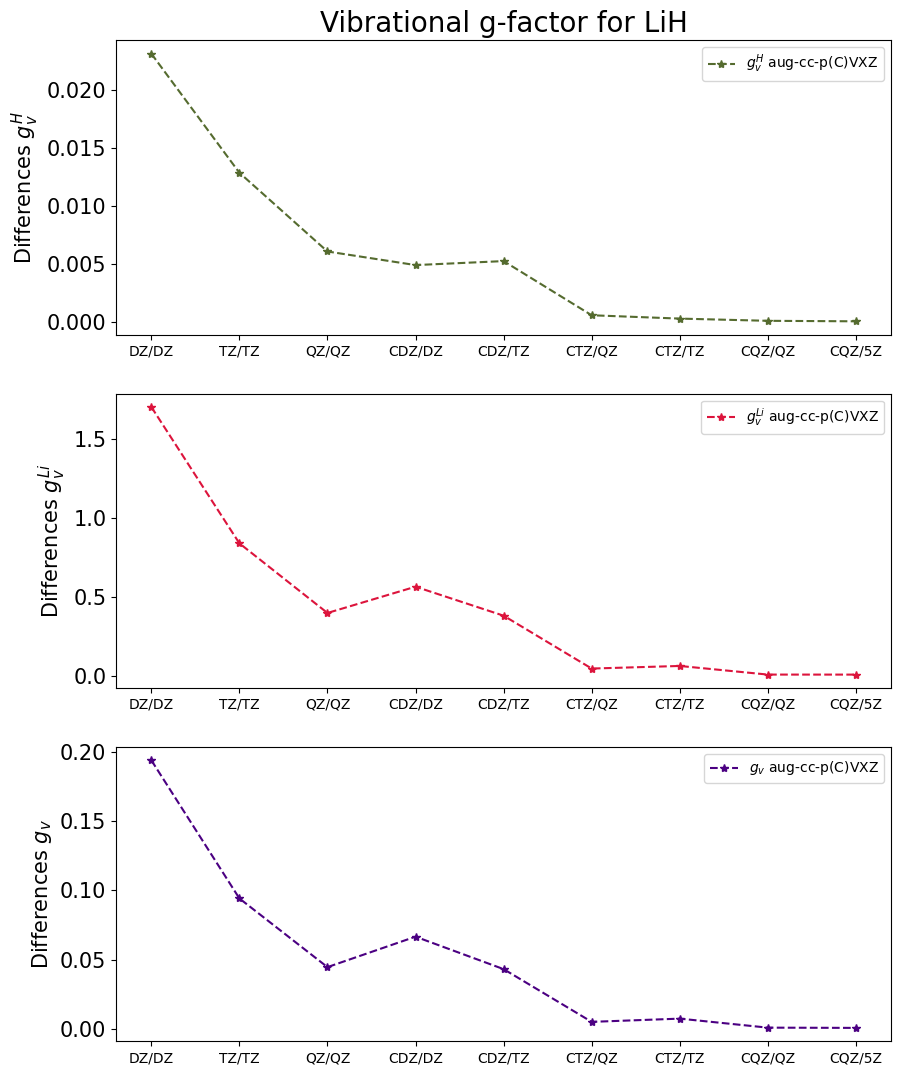} }}%
    \caption{LiH: Differences in the isotopically independent vibrational g-factors, $g_v^{Li}$ and $g_v^H$, and the total vibrational g-factor, $g_v$, calculated from eqs. \eqref{eq gvA_respons}, \eqref{eq gvB_respons} and \eqref{eq: gv} versus from eqs. \eqref{eq: gA dipole}, \eqref{eq: gB dipole} and \eqref{eq: g_V_3_cont} as a function of the basis sets for Li/H at the MCSCF level of theory. The active space is (4e, 10o).}
    \label{fig: LiH_basisset}
\end{figure}
\noindent  
For LiH different combinations of basis sets were tested, including aug-cc-p(C)VXZ with X=D-Q for Li and aug-cc-pVXZ with X=D-5 for H, see Fig. \ref{fig: LiH_basisset}. 
This analysis agrees with the tendencies observed for CO and CS, thus, the inclusion of extra core-valence functions on the basis set for Li improves dramatically the results and increasing the zeta levels also enhances the basis set sufficiency. Among $g_v^{Li}$, $g_v^{H}$ and $g_v$, the choice of basis set has the greatest impact on $g_v^{Li}$. Consequently, optimizing basis set for Li is critical to obtain reliable results.

A previous study regarding the rotational g-factors had shown that the finite basis set effect is independent of correlation.\cite{sauer1998relation} A similar trend is observed here for the vibrational g-factor when 
comparing the MCSCF and HF differences in Figs. \ref{fig: CO_basisset} and \ref{fig: CS_basisset}. 
It can thus be concluded that while the choice of basis set has a  significant influence on the precision of the results, the MCSCF and HF methods yield almost identical differences throughout. To reduce computational cost, the basis set benchmark could thus have been made solely utilizing the HF method. 

To summarize, when the zeta quality in the aug-cc-pVXZ basis sets is increased, the differences converge towards 0, thus indicating a more sufficient basis set. To further improve the accuracy of the basis set, it is preferable to include extra core-valence functions. The inclusion of a second set of diffuse functions, on the other hand, offers no additional precision and can thus safely be omitted.

\subsection{Selection of Active Space}
In this section we will discuss the effect of the size of the active space on the calculated rotational g-factors for CO and CS. As reference we will employ the experimental values.
\begin{figure}[h!]
    \centering
    \subfloat[\centering Experimental data: (0.26910 ± 0.0005)\cite{rosenblum1958isotopic}, (0.267 ± 0.003)\cite{burrus1959zeeman}, -0.26890 ± 0.00010\cite{ozier1967sign}, and 0.262 ± 0.026\cite{wang1993rotational}]{{\includegraphics[width=9cm]{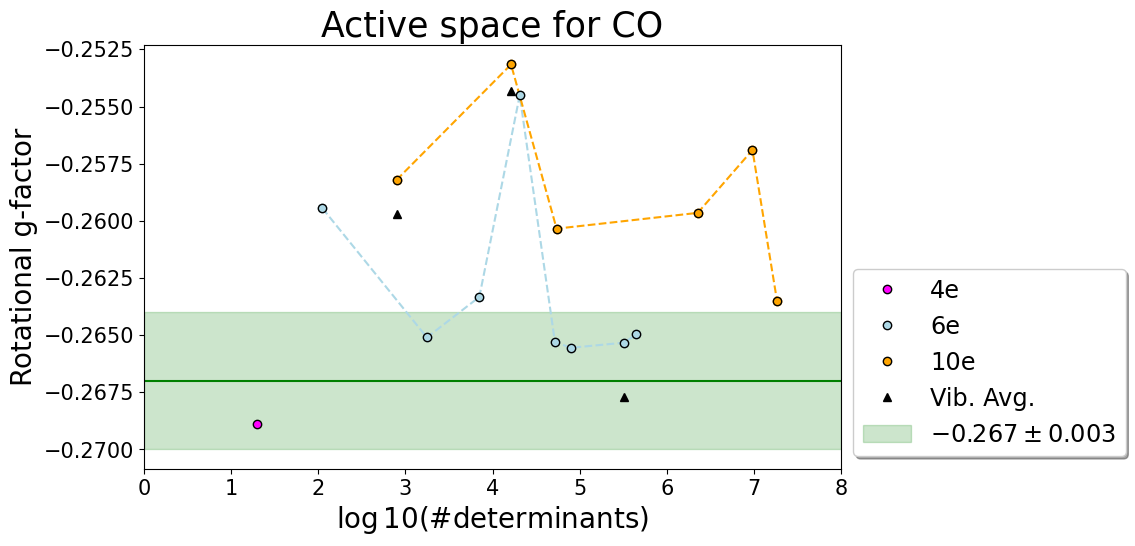} }}%
    \caption{CO: Rotational g-factor calculated with different active spaces and the aug-cc-pCV5Z basis set as function of the number of the determinants in the CI expansion.}
    \label{fig: AS CO}%
\end{figure}
\begin{figure}[h!]
    \centering
    \subfloat[\centering Experimental data: (0.269 ± 0.005)\cite{bates1968millimeter} and -0.2702 ± 0.0004\cite{mcgurk1973detection}]{{\includegraphics[width=9cm]{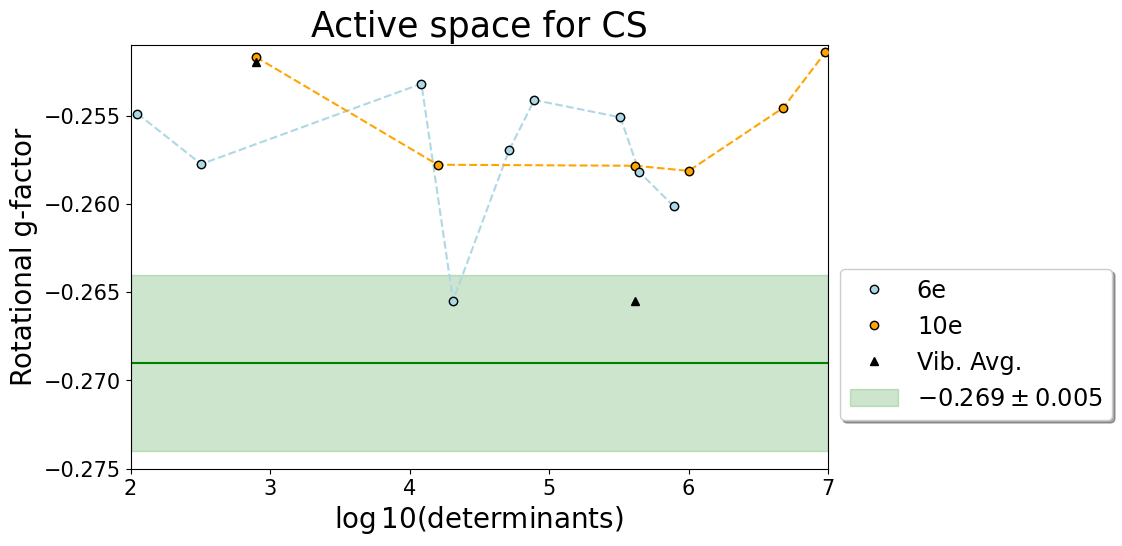} }}%
    \caption{CO: Rotational g-factor calculated with different active spaces and the aug-cc-pCV5Z basis set as function of the number of the determinants in the CI expansion.}
    \label{fig: AS CS}%
\end{figure}
Fig. \ref{fig: AS CO} for CO and \ref{fig: AS CS} for CS show the rotational g-factor calculated with the aug-cc-pCV5Z basis set and various active spaces and different number of electrons for CO and CS. The green line denotes an experimental measured value along with its reported standard error as green shaded area. 

One would expect that increasing the size of the active space would lead to a monotonic convergence towards the experimental values. Evidently, for both molecules, no such convergence is observed with the active spaces possible in this study. The computed g-factors throughout all active spaces exceed the experimentally measured values. This may be due to lack of including vibrational corrections, inefficient treatment of dynamic correlation, choice of basis set and for CS also the neglect of including relativistic effects. 
However, the chosen basis set, aug-cc-pCV5Z, has already been proven to be sufficient. 
The triangles in the plots show the vibrational averaged rotational g-factors for some of the active spaces. 
For both molecules the first vibrationally averaged rotational g-factor was calculated with an active space including the valence electrons and orbitals, \textit{i.e.} (10e, 6o) for CO and (10e,8o) for CS. Note that in the case of CS it was necessary to include two more unoccupied orbitals in order to include also the $\sigma^*$ orbital. The next vibrational averaged value was obtained with active space containing one unoccupied orbital per occupied orbital (10e,10o) for CO and (10e,13o) for CS. 
More active spaces have been further generated, by including less electrons and even more virtual orbitals. Comparing to the experimental values the (10e,10o) active space for CS proved sufficient, whereas for CO the best results were obtained when extending the active space to (6e,20o). For both CO and CS the vibrational averaged values are significantly lower, and it is expected that, if all values were averaged that the computed g-factors would approach the experimental values. 

The standard deviation of all the computed g-factors is 0.0037 and 0.0035 for CO and CS, respectively. This is in same order of magnitude as the standard error of the experiments.
Although the rotational g-factors do not converge monotonically towards the experimental value, some of the computed values fall within the respective uncertainty ranges of the experimental data. Unsurprisingly, this is observed for the experimental values with the largest uncertainties. 
The experimental g-factors shown in the figure are chosen, since they have errors close to each other for the two molecules.

\subsection{Comparison of Different Electronic Structure Theory Methods}
\begin{figure}[h!]
    \centering
   {{\includegraphics[width=7cm]{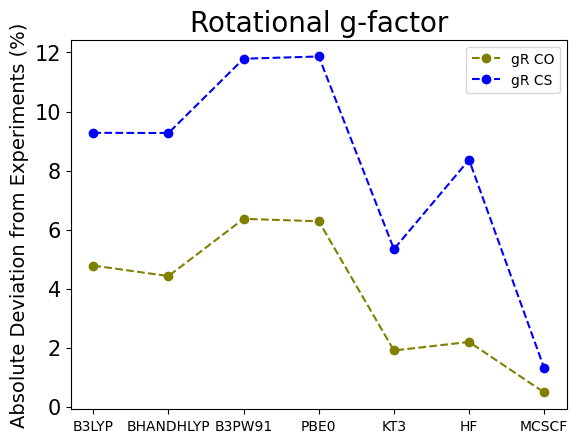} }}%
    \qquad
    \caption{Deviation of the predicted rotational g-factors for CO (green) and CS (blue) from the experimental values for different DFT functionals, HF and MCSCF. The basis set used is aug-cc-pCV5Z and the active spaces are (10e,13o) for CS and (6e,20o) for CO. The HF and MCSCF values are vibrationally averaged.}
    \label{fig:DFT_gr}
\end{figure}

\begin{figure}[h!]
    \centering
    {{\includegraphics[width=7cm]{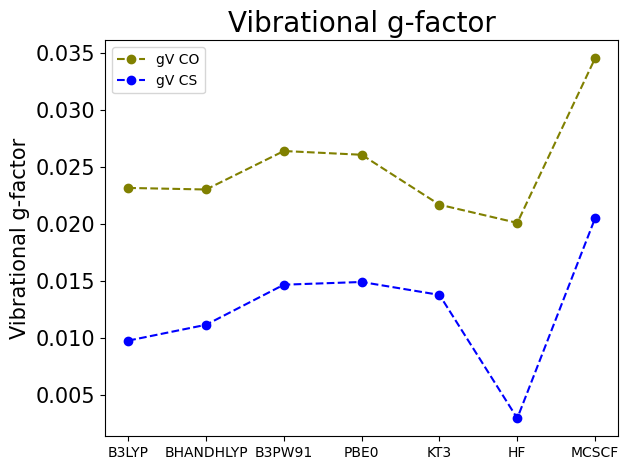} }}%
    \caption{Predicted vibrational g-factors for CO (green) and CS (blue) for different DFT functionals, HF and MCSCF. The basis set used is aug-cc-pCV5Z and the active spaces are (10e,13o) for CS and (6e,20o) for CO. The HF and MCSCF values are vibrationally averaged.}
    \label{fig:DFT_gv}
\end{figure}

Fig. \ref{fig:DFT_gr} illustrates the absolute deviation from the experimental values of the rotational g-factor for CO (green line) and CS (blue line), whereas Fig. \ref{fig:DFT_gv} shows the value of the vibrational g-factor for both molecules for different electronic structure theory methods.
The nuclear part of the g-factors for CO (0.5037) and CS (0.5038) are relatively close. The displacement of the CO curves must therefore primarily be attributed to the difference in the electronic part, that is, in the response functions.
From Fig. \ref{fig:DFT_gr} it becomes evident that MCSCF with the (6e,20o) and (10e,13o) active spaces for CO and CS, respectively, and DFT employing the KT3 method give the best results. 
When comparing the different functionals, it appears that the inclusion of Hartree-Fock exchange has little impact on the accuracy. B3LYP and BHandHLYP yield almost identical deviations even though B3LYP includes 20 \% HF exchange and BHandHLYP includes 50 \% HF exchange and B3PW91 and B3LYP have even the same amount of HF exchange. Comparing these functionals reveals that the B3PW91 functional is less reliable in these calculations than B3LYP. This indicates that the LYP correlation functional is more optimal when computing g-factors. 
The PBE0 functional (25 \% HF exchange) has only slightly higher HF exchange compared to B3LYP. However, the rotational g-factors are noticeably less accurate.  B3PW91 has same exchange as B3LYP but a different correlation. PBE0 differs in both exchange and correlation. Since B3PW91 and PBE0 have almost equal deviation, this indicates that correlation functionals are more important than the exchange functional. 
The KT3 functional is designed to be suitable for magnetic properties. It has been reported to provide accurate NMR shielding constants for light nuclei.\cite{kongsted2008accuracy}

Fig. \ref{fig:DFT_gr} supports the conclusion that computationally cheaper methods, such as HF and DFT, are inadequate compared to the MCSCF method.
The MCSCF method treats the static correlation well and can handle systems with near-degenerate electronic states. It is a crucial method for describing changes in molecular properties on varying bond lengths, where single-reference methods generally fail. Although MCSCF includes static correlations, it can be lacking in dynamic correlation, whereas DFT calculations include dynamic correlation through the exchange-correlation functional. However, Fig. \ref{fig:DFT_gr} clearly shows that MCSCF provides the most reliable results.
For the vibrational g-factor, Fig. \ref{fig:DFT_gv}, the trends of the different methods and functionals are similar for CO and CS with the exception of the MCSCF results. Nevertheless, it suggests that the methods have the same impact on the vibrational g-factor independent of the molecule.

\subsection{Geometry Dependencies}
\begin{figure}[h!]
    \centering
    \subfloat[\centering CO: Active space used is (6e,20o).]{{\includegraphics[width=7cm]{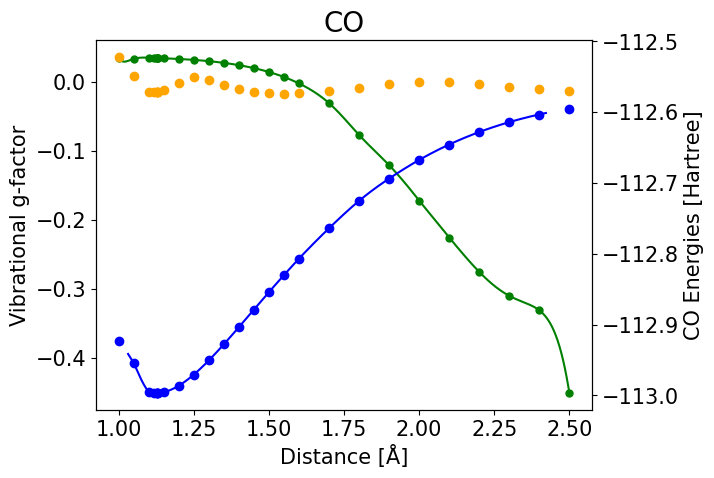} }}%
    \qquad
    \subfloat[\centering CS: Active space used is (10e,13o).]{{\includegraphics[width=7cm]{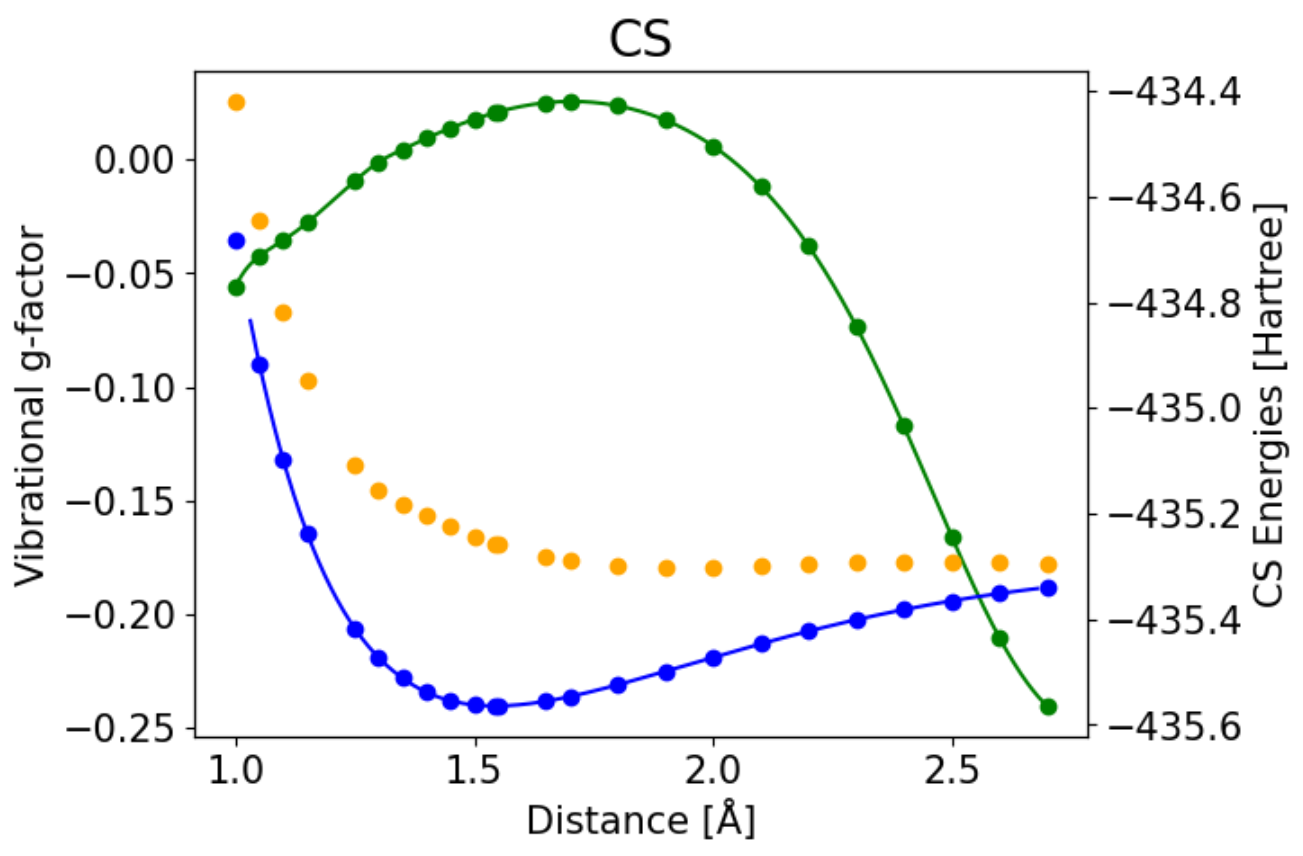} }}
    \caption{CO and CS: The dependence of the vibrational g-factor (green line), the ground state energy (blue line) and first excited state energy (orange line) on the internuclear distance calculated at the MCSCF level with the aug-cc-pCV5Z basis set.}
    \label{fig:dis CO/CS}%
\end{figure}

\begin{figure}[h!]
    \centering
    \subfloat[\centering SiO: Active space used is (10e, 8o).]{{\includegraphics[width=7cm]{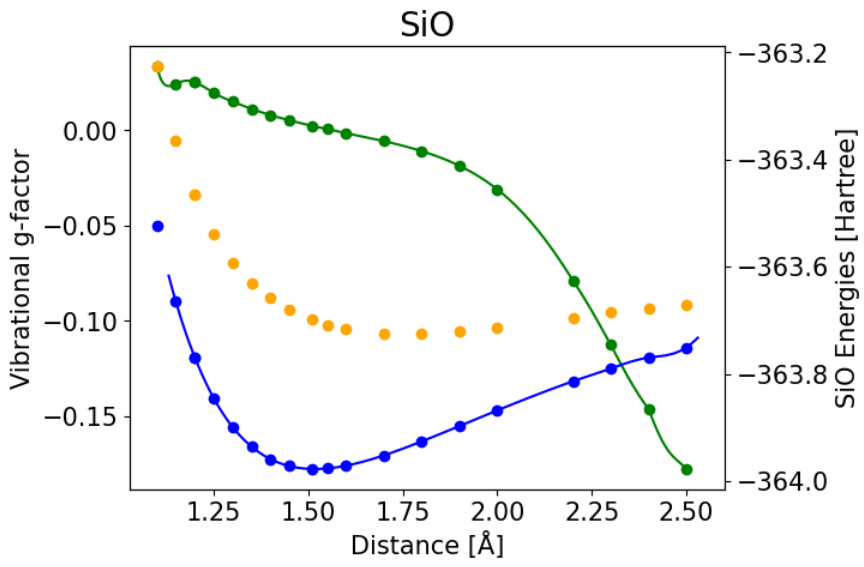} }}%
    \qquad
    \subfloat[\centering SiS: Active space used is (10e, 13o).]{{\includegraphics[width=7cm]{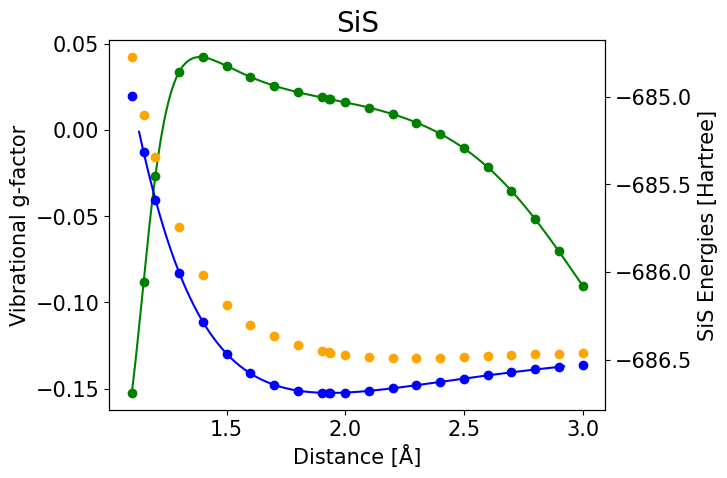} }}%
    \caption{SiO and SiS: The dependence of the vibrational g-factor (green line), the ground state energy (blue line) and first excited state energy (orange line) on the internuclear distance calculated at the MCSCF level with the aug-cc-pCV5Z basis set.}%
    \label{fig:dis SiO/SiS}%
\end{figure}

\begin{figure}[h!]
    \centering
    \subfloat[\centering LiH: The active space used is (4e, 5o).]{{\includegraphics[width=7cm]{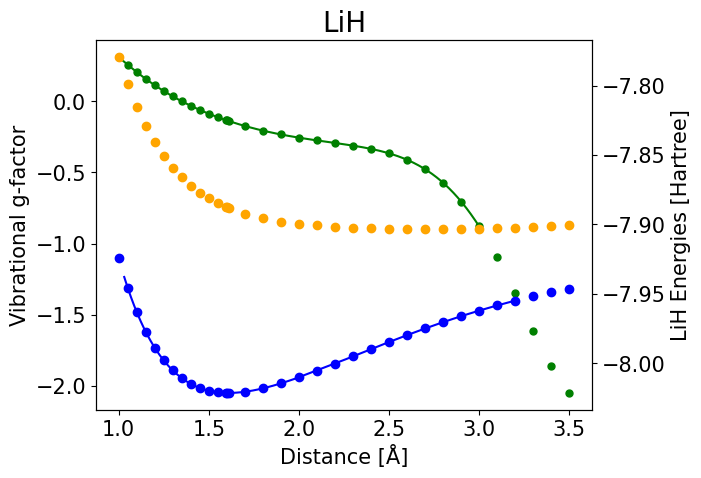} }}%
    \qquad
    \subfloat[\centering LiF: The active space used is (8e, 8o).]{{\includegraphics[width=7cm]{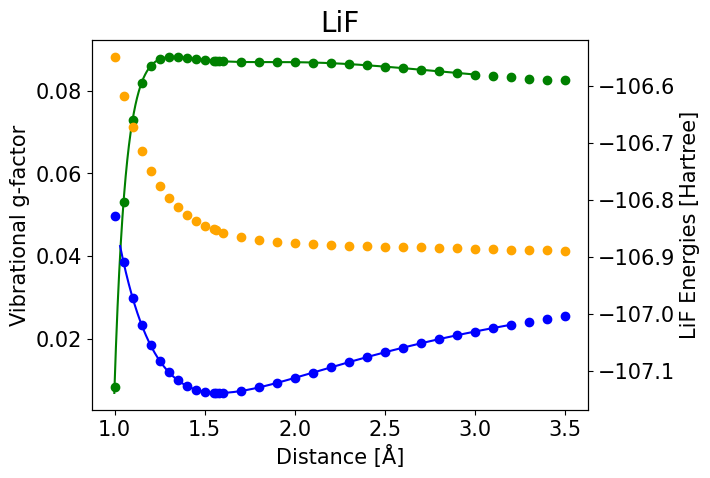} }}%
    \caption{LiH and LiF: The dependence of the vibrational g-factor (green line), the ground state energy (blue line) and first excited state energy (orange line) on the internuclear distance calculated at the MCSCF level with the aug-cc-pV5Z basis set for H, aug-cc-pCVQZ for Li and aug-cc-pCV5Z for F.}%
    \label{fig:dis LiH/LiF}%
\end{figure}

Figs. \ref{fig:dis CO/CS}, \ref{fig:dis SiO/SiS}, and \ref{fig:dis LiH/LiF}, present the total vibrational g-factor (green),
the ground state MCSCF energy (blue) and the first excited state MCSCF energy (orange) at different internuclear distances for CO, CS, SiO, SiS, LiH and LiF. The excited energy curves are chosen to be of the same symmetry as the ground-state and are chosen based on states where the non-adiabatic coupling elements of the numerator in the response functions, eq. \eqref{eq: gv}, are of significance. 

In Fig. \ref{fig:dis CO/CS} the curves for CO and CS are presented. Corresponding curves for CO obtained with the valence active space or the (10e,10o) active space would exhibit the same tendencies as observed for (6e,20o) active space shown here. 

Cooper and Kirkby\cite{cooper1987theoretical, kirby1989theoretical} reported potential energy curves for five low-lying adiabatic states of CO, using MCSCF with all valence orbitals in the active space. The excited energy curve of CO, shown in Fig. \ref{fig:dis CO/CS} (a), is similar to the reported $2^1\Sigma^+$ energy curve of Cooper and Kirkby. Similarly, Sun and Shi report potential energy curves for CS.\cite{sun2020radiative} The curves are calculated employing the CASSCF method, followed by an icMRCI approach. The aug-cc-pV5Z and aug-cc-pV6Z basis sets were used for carbon and aug-cc-pV(5 + d)Z and aug-cc-pV(6 + d)Z for sulfur. The $\Sigma^+$ curve aligns well with the reported curve. This confirms that the chosen energy curve is of correct symmetry.
As the internuclear distance increases, the energies of the ground and first excited states converge toward degeneracy. This directly influences the electronic contribution to the vibrational g-factor, as the denominator in the response function, eq. \eqref{eq: gv}, also decreases. Consequently, the electronic contribution to the vibrational g-factor increases. Since the sign of the total vibrational g-factor is negative, the relative magnitude of the electronic part must exceed that of the nuclear contribution.
For CO, the vibrational g-factor remains almost constant within the region of 1.00-1.25 Å and subsequently decreases. This behavior can be partially attributed to the nearly parallel shifts of the energy curve minima in both the ground and excited states. As the energies approach each other, the vibrational g-factor is influenced accordingly.
In contrast, CS shows a different trend. In the region around 1.00-1.70 Å the energy differences and accordingly the g-factor increase. Beyond that region, the curvature of the vibrational g-factor resembles that of CO.

Fig. \ref{fig:dis SiO/SiS} shows the curves for SiO and SiS.
The potential energy surfaces for SiO have previously been studied by Langhoff and Arnold using the self-consistent field method combined with the configuration interaction method.\cite{langhoff1979theoretical} The ground state energy curve corresponds well to our results and the $E^1\Sigma^+$ curve agrees with the reported excited energy.
Wu and co-workers\cite{li2024spectroscopy} evaluated very recently the potential energy surface of SiS at the MRCI+Q/aug-cc-pwCV5Z-DK level of theory.
Regarding the vibrational g-factors the connection between energy states and vibrational g-factor aligns with the observations for CO and CS. 
For SiS, the sign of the vibrational g-factor at 1.00 Å is negative, thus the electronic part dominates. This is consistent with the observation that the energy curves are relatively close in this region. As long as an increase in internuclear distance provides larger separation between the energy states the total g-factor increases. In contrast, as the energy levels approach each other the total g-factor decreases again. This trend is observed for both SiO and SiS.

\begin{table}[h!] 
\small
    \caption{Vibrational and rotational g-factors computed at experimentally measured equilibrium and computed equilibrium using basis set aug-cc-pCV5Z for C, O, S, Si F, aug-cc-pCVQZ for Li and aug-cc-pV5Z for H.}
    \label{tab: vibavg}
    \begin{tabular}{llrrrrrr} 
    \hline 
    AB& Method &
    $g_v^{\text{r}_\text{exp}}$&
    $g_v^{\text{r}_\text{calc}}$&
    $\left<g_v\right>$&
    $g_{r}^{\text{r}_\text{exp}}$&
    $g_{r}^{\text{r}_\text{calc}}$&
    $\left<g_{r}\right>$\\
    \hline
CO  & MCSCF (10e,6o)  & 0.0349& 0.0348 & 0.0347& -0.2582 & -0.2588  &-0.2597 \\
    & MCSCF (10e,10o) & 0.0352& 0.0352 & 0.0351& -0.2532 & -0.2533  &-0.2543 \\
    & MCSCF (6e,20o)  & 0.0349& 0.0349 & 0.0345& -0.2653 & -0.2651  &-0.2677 \\
    &    HF           & 0.0188& 0.0203 & 0.0201& -0.2805 & -0.2734  &-0.2749 \\
CS  & MCSCF (10e,8o)  & 0.0201& 0.0206 & 0.0206& -0.2517 & -0.2518  &-0.2520\\
    & MCSCF (10e,13o) & 0.0200& 0.0205 & 0.0205& -0.2655 & -0.2655  &-0.2656 \\
    &    HF           & 0.0040& 0.0030 & 0.0030& -0.2958 & -0.2980  &-0.2915\\
SiO & MCSCF (10e,8o)  & 0.0024& 0.0021 & 0.0019& -0.1420 & -0.1421  &-0.1424\\
    &    HF           &-0.0083&-0.0088 &-0.0059& -0.1529 & -0.1534  &-0.1518\\
SiS & MCSCF (10e,13o) & 0.0182& 0.0180 & 0.0179& -0.0856 & -0.0858  &-0.0860\\
    &    HF           & 0.0125& 0.0133 & 0.0132& -0.0918 & -0.0912  &-0.0915\\
LiH & MCSCF (4e,5o)   &-0.1314&-0.1380 &-0.1429& -0.6638 & -0.6510  &-0.6447\\
    &    HF           &-0.1699&-0.1751 &-0.1826& -0.6966 & -0.6881  &-0.6845\\
LiF & MCSCF (8e,8o)   & 0.0872& 0.0872 & 0.0872&  0.0683 &  0.0700  &0.0701\\
    &    HF           & 0.0892& 0.0893 & 0.0893&  0.0721 &  0.0712  &0.0714\\
    \hline
    \end{tabular}
    \footnotesize{
     Experimental values: CO: $g_r=(0.26910 \pm 0.0005)$\cite{rosenblum1958isotopic}, $(0.267 \pm 0.003)\cite{burrus1959zeeman}, -0.26890 \pm 0.00010$\cite{ozier1967sign}, $0.262 ± 0.026\cite{wang1993rotational}$ 
     CS:$g_r= (0.269 \pm 0.005)$\cite{bates1968millimeter}, $-0.2702 \pm 0.0004$\cite{mcgurk1973detection}, 
    SiO: $g_r = -0.15359 \pm 0.00012$\cite{davis1974magnetic}, 
    SiS: $g_r =-0.09097 \pm 0.000065$\cite{honerjager1973gj}, 
     LiH $g_r = -(0.654 \pm 0.007)$\cite{
    lawrence1963rotational}, $-0.65842 \pm 0.00017$\cite{freeman1975molecular}, 
    LiF: $g_r = (0.0642 \pm 0.0004)$\cite{russell1958magnetic}, $(0.07367 \pm 0.00050)$\cite{mehran1966rotational}. \\
    Previously computed values:\\ 
    CO: $g_r^{\text{SCF}}=-0.2800$,\cite{kjaer2009relation}
    $g_r^{\text{MCSCF}}=-0.2577$,\cite{kjaer2009relation} 
    $\left<g_r\right>^{\text{MCSCF}}=-0.259$,\cite{koput2024toward} 
    $g_r^{\text{CCSD(T)}}=-0.2678$,\cite{mag09-jcp131-144104} 
    $g_v^{\text{SCF}}=0.0193$,\cite{kjaer2009relation} 
    $g_v^{\text{MCSCF}}=0.0355$,\cite{kjaer2009relation}
    $\left<g_v\right>^{\text{MCSCF}}=-0.046$,\cite{koput2024toward} \\ 
    SiO: $g_r^{\text{SCF}}=-0.1522$,\cite{kjaer2009relation} 
    $g_r^{\text{MCSCF}}=-0.1413$,\cite{kjaer2009relation} 
    $g_v^{\text{SCF}}=-0.0074$,\cite{kjaer2009relation} 
    $g_v^{\text{MCSCF}}=0.0033$,\cite{kjaer2009relation}\\ 
    LiH: $g_r^{\text{SCF}}=-0.6962$,\cite{kjaer2009relation} 
    $g_r^{\text{MCSCF}}=-0.6613$,\cite{kjaer2009relation} 
    $g_r^{\text{MCSCF}}=-0.6647$,\cite{sauer2011calculated} 
    $g_r^{\text{CCSD(T)}}=-0.6638$,\cite{mag09-jcp131-144104} 
    $g_v^{\text{SCF}}=-0.1689$,\cite{kjaer2009relation} 
    $g_v^{\text{MCSCF}}=-0.1320$,\cite{kjaer2009relation} 
    $g_v^{\text{MCSCF}}=-0.1366$,\cite{sauer2011calculated}\\ 
    LiF:$g_r^{\text{SCF}}=0.0729$,\cite{kjaer2009relation} 
    $g_r^{\text{MCSCF}}=0.0691$,\cite{kjaer2009relation} 
    $g_r^{\text{CCSD(T)}}=-0.0677$,\cite{mag09-jcp131-144104} 
    $g_v^{\text{SCF}}=0.0894$,\cite{kjaer2009relation} 
    $g_v^{\text{MCSCF}}=0.0875$.\cite{kjaer2009relation}
    }
\end{table}
The energy curves for LiH and LiF have been discussed by Stwalley and Zemke\cite{stwalley1993spectroscopy}, and Botter, Kooter and Mulder\cite{botter1975ab}, respectively. Comparing their reported energy curves to our curves, see Fig. \ref{fig:dis LiH/LiF}, confirms that the ground state and first excited state agree and are of same symmetry. The patterns for LiH and LiF in Fig. \ref{fig:dis LiH/LiF} empathize the trends already observed for the other molecules. Both molecules contain a longer region with an almost constant g-factor, as the energy states merely approach each other gradually. 

The vibrational g-factors are expected to be zero for $R$=0 and for $R\rightarrow \infty$. For $R\rightarrow \infty$ we expect no interaction between the two atoms, thus the electrons are confined to one of the atoms; consequently, no correction factor is observed. The sign of the vibrational g-factors depends solely on the relative magnitudes of the nuclear and electronic contributions to the g-factor, since the nuclear contribution is inevitably positive and the electronic contribution is negative.
With the exception of LiH, the g-factors are positive in the equilibrium geometry, and hence the nuclear contribution is more significant. However, as the distance increases, the g-factor decreases and becomes negative, thus the relative magnitude of the electronic contribution increases.

\subsection{Equilibrium Geometry and Vibrationally Averaged Values}

Tab. \ref{tab: vibavg} finally shows the rotational and vibrational g-factors computed at the experimental and the optimized equilibrium distances as given in Tab. \ref{tab: Dis} along with the vibrational averaged rotational and vibrational g-factors at the MCSCF and HF level. 

At the MCSCF level the difference between values calculated at the experimental and computed equilibrium bond lengths is close for all molecules. The differences vary on the fourth decimal for all g-factors except the rotational g-factor of LiH and LiF, where it already varies on the second decimal. As expected, at HF level the experimental and computed equilibrium bond length vary more. Since HF usually underestimates bond lengths, see Tab. \ref{tab: Dis}, the computed HF equilibrium bond distance differs more from the experimental equilibrium bond distance compared to MCSCF and thus a greater difference in g-factors is observed. 

Except for LiH, all of the vibrational averaged values are close to those of the vibrational g-factor at the computed equilibrium level. For LiH the vibrational averaging has a larger impact; this is expected because LiH contains hydrogen that is a light molecule and is thus expected to vibrate more. Generally, it is observed that the MCSCF values are closer to the vibrational averaged values compared to HF, and often the g-factors calculated at the computed equilibrium are closer or equally close to the vibrational averaged values compared to the g-factors calculated at experimental equilibrium. Comparing MCSCF and HF values clearly shows that the relative magnitude is larger at the HF level. In the article by Kjær and Sauer\cite{kjaer2009relation} it is discussed that HF calculations in general underestimate the rotational g-factor. We observe that tendency slightly, however, most of the rotational g-factors are improved through vibrational averaging when compared to experimental values. 

For LiH, the active space contains 4 electrons in 5 orbitals. This active space is thus significantly smaller than the one reported in the article by Sauer, Oddershede and Ogilvie.\cite{sauer2011calculated} The vibrational g-factor at different geometries differs slightly, but the overall tendency is preserved. At equilibrium distance the reported g-factors are $g_v=-0.1366$ and $g_r=-0.6647$, whereas ours become $g_v=-0.1314$ and $g_r=-0.6638$. Upon comparing the equilibrium rotational g-factor with experimentally determined values our g-factor agrees slightly better with the experimental values. The computed results agree well with the g-factors reported in the article by Kjær and Sauer,  $g_v=-0.1320$ and $g_r=-0.6613$. 

For CO, SiO and LiF the g-factors are close to the previously estimated values of Kjær and Sauer, however, the ones reported here tend to be slightly bigger. The active spaces by Kjær and Sauer are (10e,8o) for CO, (8e, 6o) for LiF and (10e, 8o) for SiO. The active space of SiO equals the one we have used. For LiF we have used a larger active space, and for CO we have included more orbitals but fewer electrons. The basis sets we have used are much larger than the basis sets used in the previous article. 

For SiO, they also report a negative sign of the vibrational g-factor at the HF level and a positive sign at the MCSCF level. The HF value differs somewhat, since they report $g_v=-0.0074$. The vibrational g-factor at MCSCF level also differs slightly for SiO compared to the other molecules, as it is $g_v=0.0033$. The rotational g-factors are all more similar.

Koput\cite{koput2024toward} predicted the vibrational averaged rotational and vibrational g-factors for CO to be $\left<g_r\right> =-0.259$ and $\left<g_v\right> =0.046$, respectively. These values are close, yet slightly higher than the once obtained here.

Lutnæs and coworkers have also reported rotational g-factors for CO, LiH and LiF using the CCSD(T) method with different variations of basis sets.\cite{mag09-jcp131-144104} Using the aug-cc-pCVQZ basis set they obtain $g_r=-0.2678$ for CO, $g_r=-0.6638$ for LiH and $g_r=0.0677$ for LiF. These results are close to the ones we have obtained using MCSCF.

\section{Conclusion}
In this article, we examined the most suitable methods and basis sets for computing g-factors, including HF, DFT, and MCSCF. Various exchange-correlation functionals, specifically B3LYP, BHandHLYP, B3PW91, PBE0, and KT3 were analyzed in DFT calculations. Additionally, different active spaces were explored to improve the MCSCF calculations.
The computed rotational g-factors were compared to experimentally measured values, as an indicator of the accuracy of the calculations.
The results revealed that MCSCF provided the most accurate results, followed by DFT employing the KT3 functional. 
For CO and CS, the basis set benchmark suggested aug-cc-pCV5Z as the most appropriate basis set, while for LiH, the combination of aug-cc-pCVQZ for Li and aug-cc-pV5Z for H yielded the most accurate results.

For CO, CS, SiO, SiS, LiH and LiF the relationship between the energy difference of the ground state and first excited state and the vibrational g-factor was further explored. Additionally, it was investigated how a change in the internuclear distance influences the vibrational g-factor. It was found that as the ground state and first excited state energy levels approach each other, the vibrational g-factor becomes increasingly negative, thus the magnitude of the electronic contribution exceeds the magnitude of the nuclear contribution.

Eq. \eqref{eq: gv} shows how the electronic contribution depends on the energy differences between the ground state and excited states. Ideally, all energy levels contribute to the sum. However, we have only investigated how the difference between the ground state and first excited state influences the vibrational g-factor. Nevertheless, it is assumed that the first excited state is the biggest contributor to the sum. Essentially, the g-factor will follow the same tendency as if all levels were included. 
Although also evaluating the numerator of the response function would yield more precise insights, as multiple non-adiabatic coupling elements could contribute significantly to the numerator. Only plotting the non-adiabatic coupling elements that correspond to the first excited state would not provide the correct indication of the numerator's contribution to the response function. 
\\
Furthermore, we computed the vibrationally averaged rotational and vibrational g-factors for CO, CS, SiO, SiS, LiH, and LiF. It appears that vibrational averaging improves the agreement between computed g-factors and experimental data.



\section*{Data Availability}

The data that support the findings of this study are available from the corresponding authors upon reasonable request.

\providecommand{\latin}[1]{#1}
\makeatletter
\providecommand{\doi}
  {\begingroup\let\do\@makeother\dospecials
  \catcode`\{=1 \catcode`\}=2 \doi@aux}
\providecommand{\doi@aux}[1]{\endgroup\texttt{#1}}
\makeatother
\providecommand*\mcitethebibliography{\thebibliography}
\csname @ifundefined\endcsname{endmcitethebibliography}  {\let\endmcitethebibliography\endthebibliography}{}

\end{document}